\documentclass[a4paper, oneside, twocolumn, notitlepage, 10pt]{extarticle_ecoc}
\usepackage{ecoc}
\usepackage{stfloats}
\usepackage{upgreek}
\begin{document}
\selectlanguage{english}    


\title{1.23-Tb/s per Wavelength Single-Waveguide On-Chip Optical Interconnect Enabled by Mode-division Multiplexing}%


\author{
    Hanzi Huang\textsuperscript{(1)},
    Yetian Huang\textsuperscript{(1)},
    Yu He\textsuperscript{(2)},
    Haoshuo Chen\textsuperscript{(3)},
    Yong Zhang\textsuperscript{(2)},
    Qianwu Zhang\textsuperscript{(1)},\\
    Nicolas K. Fontaine\textsuperscript{(3)},
    Roland Ryf\textsuperscript{(3)},
    Yingxiong Song\textsuperscript{(1)},
    and Yikai Su\textsuperscript{(2*)}
}

\maketitle                  


\begin{strip}
 \begin{author_descr}

   \textsuperscript{(1)} Key Laboratory of Specialty Fiber Optics and Optical Access Networks, Joint International Research Laboratory of Specialty Fiber Optics and Advanced Communication, Shanghai University, China
   
   \textsuperscript{(2)} State Key Laboratory of Advanced Optical Communication Systems and Networks Department of Electronic Engineering, Shanghai Jiao Tong University, Shanghai 200240, China  *\textcolor{blue}{\uline{yikaisu@sjtu.edu.cn}}

   \textsuperscript{(3)} Nokia Bell Labs, 791 Holmdel Rd., Holmdel, New Jersey 07733, USA

 \end{author_descr}
\end{strip}

\setstretch{1.1}


\begin{strip}
  \begin{ecoc_abstract}
    We experimentally demonstrate a record net capacity per wavelength of 1.23~Tb/s over a single silicon-on-insulator (SOI) multimode waveguide for optical interconnects employing on-chip mode-division multiplexing and 11$\times$11 multiple-in-multiple-out (MIMO) digital signal processing.
  \end{ecoc_abstract}
\end{strip}


\section{Introduction}
Optical interconnects have been discussed as a successor to electrical interconnects in data-center and high-performance computing systems, especially for network switches, to sustain the fast growing data rate while reducing power consumption~\cite{Winzer:18}.
Mode-division multiplexing (MDM) over multimode fiber (MMF) has been explored to scale up the system capacity by leveraging spatial mode parallelism, where MDM transmission based on 45 spatial modes has been demonstrated~\cite{8535536}.
However, on-chip MDM transmission~\cite{22,33} has yet to demonstrate employing more than 4 waveguide modes, which is mainly limited by modal crosstalk.
Multiple-in-multiple-output (MIMO) digital signal processing (DSP) can tackle mode crosstalk and has been demonstrated in real-time implementations~\cite{7323761,Beppu:s}

In this work, we demonstrate on-chip 11 transverse electric (TE)-mode MDM optical interconnects incorporating advanced modulation formats such as 16-ary quadrature amplitude modulation (16-QAM) and MIMO-based DSP on silicon-on-insulator (SOI) platform.
Ten directional couplers (DCs) and subwavelength grating (SWG) structure based mode converters for TE$_1$ to TE$_{10}$ waveguide modes are employed as the mode multiplexer and demultiplexer for 11 TE-mode on-chip transmission.
A time division multiplexing (TDM) scheme \cite{vanUden:14,Huang:20} is applied for MDM signal detection.
We obtain bit-error rate (BER) below 7\% forward error correction (FEC) threshold using quadrature phase shift keying (QPSK) and 16-QAM for all the modes over 7 wavelengths distributed from 1530~nm to 1560~nm.
We achieve a record net capacity of 1.23 Tb/s per wavelength using a single optical waveguide, which can offer a potential spectral efficiency of 37.2~bit/s/Hz using wavelength division multiplexing (WDM).



\begin{figure*}[t]
   \centering
    \includegraphics[width=\hsize]{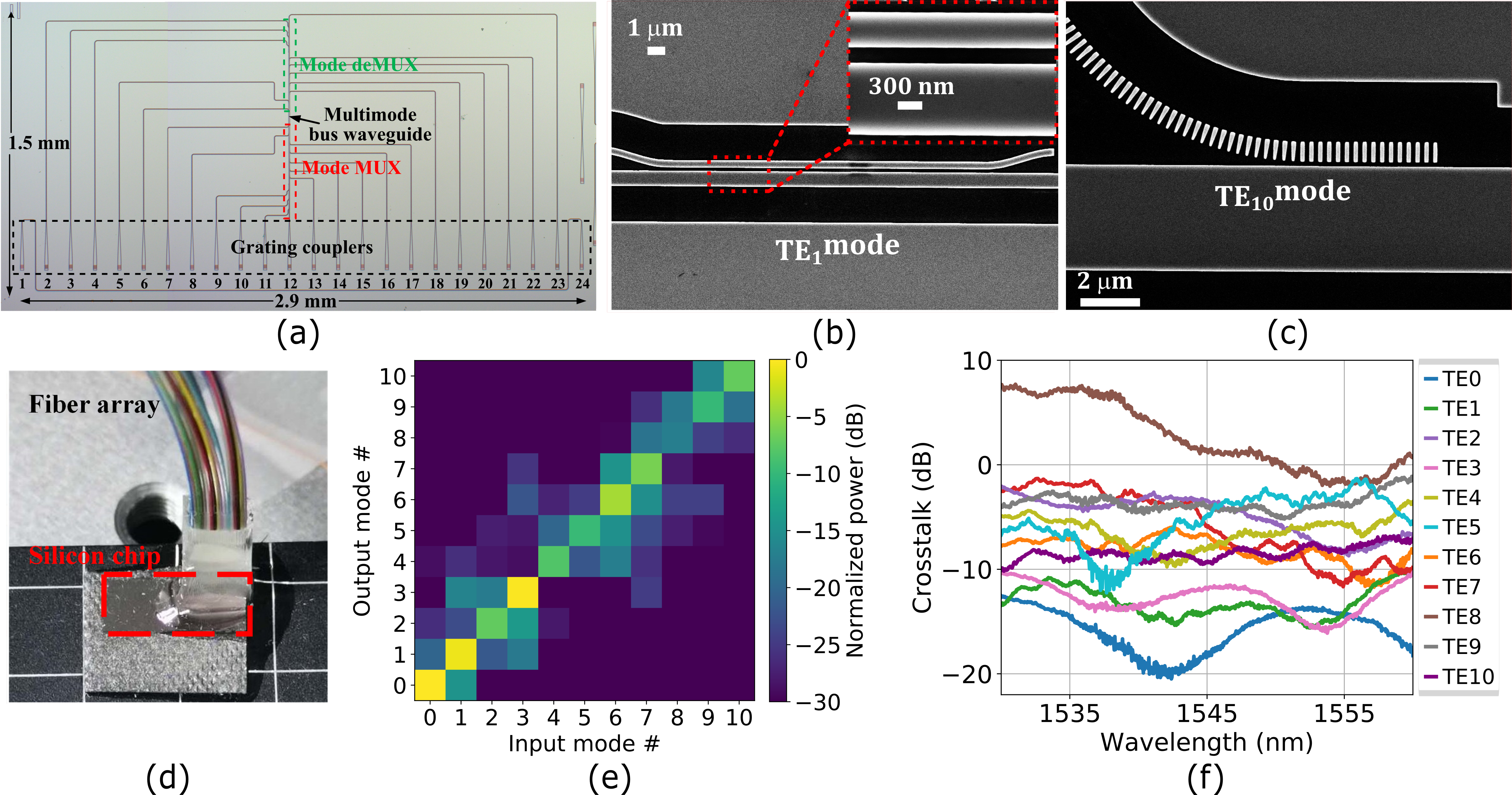}
    \setlength{\abovecaptionskip}{-0.2cm}
    \setlength{\belowcaptionskip}{-0.2cm}
    \caption{(a)~Microscope photo of the fabricated mode (de)multiplexing chip, 
    scanning electron microscope (SEM) images of the directional couplers (DCs) of (b)~TE$_1$ and (c)~TE$_{10}$, 
    (d)~photo of the packaged MDM silicon circuit, 
    (e)~measured power transfer matrix at 1550~nm, 
    (f)~measured crosstalk spectra for different modes.}
    \label{figure1}
\end{figure*}

\section{Silicon-on-insulator 11-TE-mode MDM circuit}
Figure \ref{figure1}(a) shows the microscope photo of the fabricated on-chip MDM device \cite{8514018}.
Photonic integrated MDM device was fabricated on a SOI wafer by electron-beam lithography and inductively coupled plasma (ICP) etching.
Three assymmetric DCs for TE$_1$-TE$_3$ and seven SWG-based DCs for TE$_4$-TE$_{10}$ are employed to selectively couple the injected light to different high-order modes into the multimode bus waveguide.
Scanning electron microscope (SEM) images of the DCs used for multiplexing TE$1$ and TE$_{10}$ are shown in Fig. \ref{figure1}(b) and (c), respectively.
A 24-channel fiber array is mounted on the chip using ultraviolet (UV) light curable adhesive.
Figure \ref{figure1}(d) provides the photo of packaged silicon chip and fiber array.
The measured power transfer matrix at 1550~nm by a spectrometer is shown in Fig. \ref{figure1}(e).
The measured crosstalk spectra for different modes is given in Fig. \ref{figure1}(f), where TE$_8$ experiences the strongest crosstalk of roughly 7~dB around 1530 to 1535~nm, which is mainly caused by fabrication imperfections and significantly degrades transmission performance without using joint signal processing such as MIMO-based DSP at the receiver.

\section{Experimental setup}
\begin{figure*}[b!]
   \centering
    \includegraphics[width=\hsize]{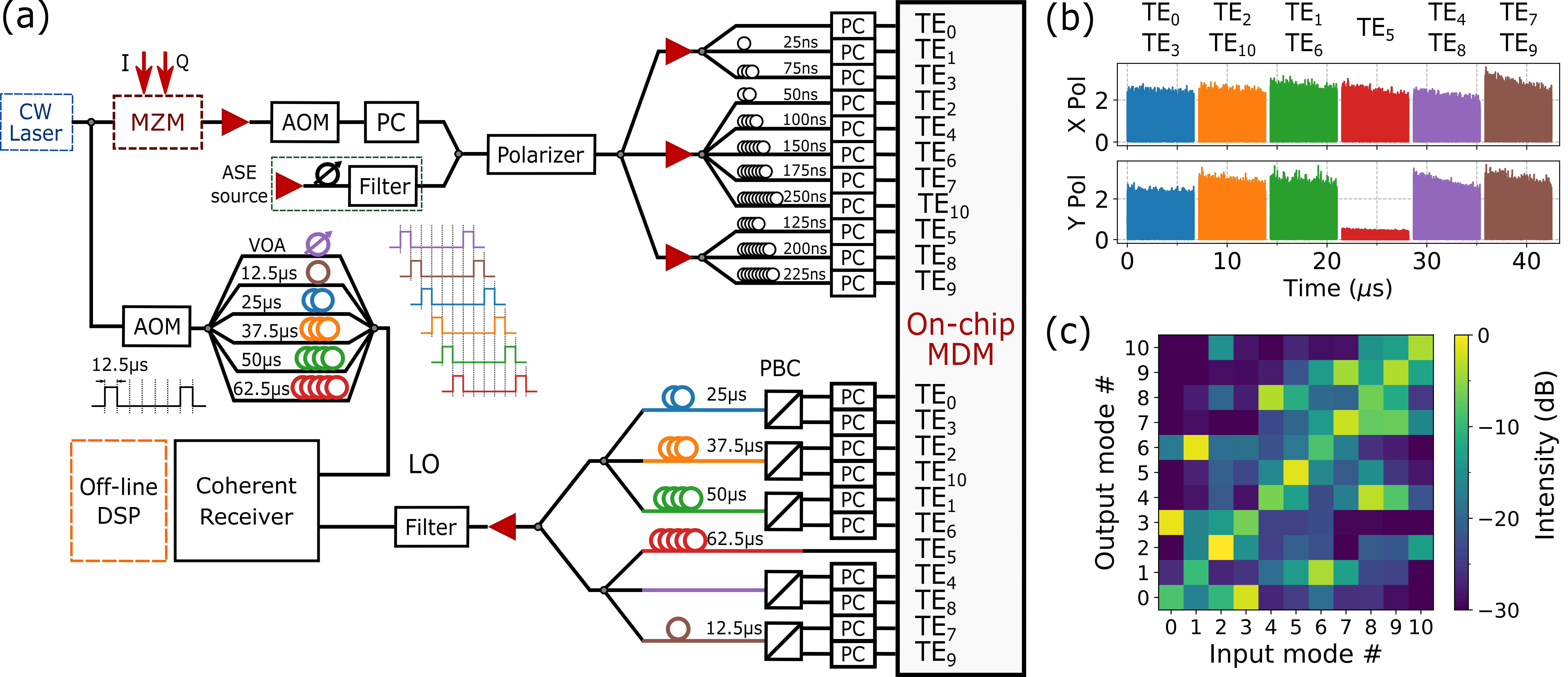}
    \setlength{\abovecaptionskip}{-0.2cm}
    \setlength{\belowcaptionskip}{-0.2cm}
    \caption{(a)~Setup for MIMO based on-chip 11-mode transmission,
    (b)~amplitudes of the six temporally stitched waveforms for 11 modes (the two polarization states carry two TE modes, except TE$_5$),
    (c)~estimated intensity transfer matrix showing the systematic mode coupling between 11 modes at 1550 nm.
    CW: continuous wave, MZM: Mach-Zehnder modulator, AOM:~acousto-optic modulator, PC: polarization controller, ASE: amplified spontaneous emission, PBC: polarization beam combiner, LO: local oscillator.}
    \label{figure2}
\end{figure*}

\begin{figure*}[t]
   \centering
    \includegraphics[width=\hsize]{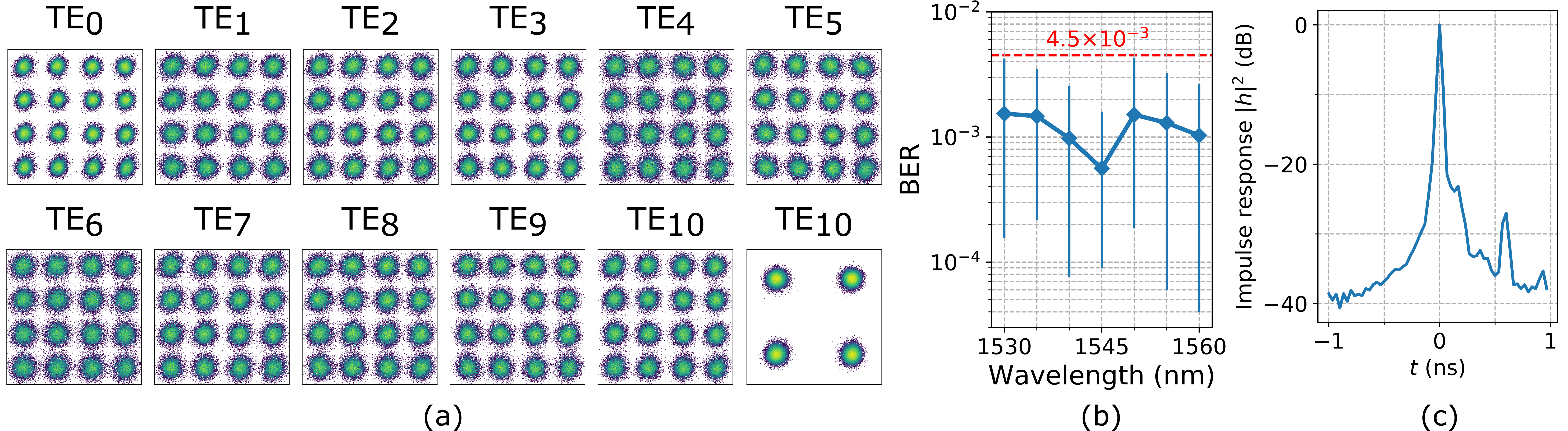}
    \setlength{\abovecaptionskip}{-0.2cm}
    \setlength{\belowcaptionskip}{-0.2cm}
    \caption{(a)~Recovered constellation of 30-GBaud 16-QAM signals for 11 modes, and QPSK signal for TE$_{10}$,
    (b)~averaged bit-error rate (BER) for 16-QAM over all modes as a function of wavelength, where error bars represent the best and the worst mode tributaries,
    (c)~normalized intensity impulse response obtained from the channel estimation at 1550 nm.}
    \label{figure3}
\end{figure*}

The experimental setup of 11-TE-mode transmission over the SOI circuit is shown in Fig. \ref{figure2}(a).
The transmitter consists of a tunable external cavity laser (ECL) with 100-kHz linewidth as light source and is modulated by an Inphase and Quadrature Mach-Zehnder modulator (IQ-MZM) driven by a high-speed digital-analog converter (DAC) operated at 60 GSamples/s with a pseudo random pattern of length 2$^{17}$, generating a Nyquist-shaped QPSK or 16-QAM signal with a roll-off factor of 0.01.
The signal is amplified and then time gated using an acousto-optic modulator (AOM) operating at 16.7\% duty cycle and a period of 75~$\upmu$s.
The time-gated signal is combined with a filtered wide-band amplified spontaneous emission (ASE) source to emulate WDM transmission before being passed through a polarizer, split and boosted by 3 erbium-doped fiber amplifiers (EDFAs).
The ASE power is large enough to saturate the 3 EDFAs to avoid transient due to signal gating.
The 3 EDFAs are with different output powers in order to balance the coupling and insertion losses of different modes.
The 3 output paths are further split to produce 11 copies of the modulated signal.
Each copy is delay-decorrelated with a relative delay of 25~ns and polarization-aligned before coupling into the MDM chip.
The input signals are then spatially multiplexed and transmitted over a multimode bus waveguide with a 70-$\upmu$m length before being demultiplexed and coupled into the output fiber array.

Five polarization beam combiners are used to combine each two TE modes into the two polarization states at one time slot, except mode TE$_5$, which can help to reduce the required number of time slots and avoid using additional sampling scopes.
The polarization multiplexed signals are sent into respective delay fiber spools with lengths of 2.5, 5, 7.5, 10 and 12.5~km and recombined together after getting separated in time.
The temporal combined signals are amplified and the out-of-band ASE is filtered out before being captured by a polarization-diverse coherent receiver.
The local oscillator (LO) path is also gated and delayed in a similar way as the signal to eliminate the impact of the LO phase variations on system performance.
The detected electrical signal is captured by a digital storage oscilloscope (DSO) with sampling rate of 40-GSa/s and processed off-line.
The received TDM signal is reconstructed and parallelized as 12 waveforms for the 11 received TE modes.
After realigning the waveform segments, the mode-multiplexed signals are able to be recovered using a frequency-domain 11$\times$11 MIMO equalizer with 80 half symbol-duration-spaced taps employing data-aided least-mean-square (LMS) algorithm.

\section{Transmission results}
The amplitudes of the six temporally stitched waveforms for the 11 modes is given in Fig. \ref{figure2}(b), where small temporal gaps caused by inaccurate length of the delay fiber spool are removed.
Figure \ref{figure2}(c) shows the estimated system intensity transfer matrix between the 11 modes measured at 1550~nm.
Note that due to the power coupling after the polarization beam combiners, the system transfer matrix is no longer diagonal as the one measured directly for the MDM chip as shown in Fig.~\ref{figure1}(e).
Figure \ref{figure3}(a) shows the recovered constellation of 30-GBaud 16-QAM signals for all the modes and QPSK signal constellation for TE$_{10}$ at 1545~nm.
BER $<7\times10^{-6}$ QPSK transmission can be achieved for each modes at the measured 7 wavelengths with no bit error observed in the whole received frames.
The averaged BER results for 16-QAM over 11 modes versus wavelength is shown in Fig. \ref{figure3}(b), where the best and worst spatial tributaries are represented as error bars.
The averaged BERs remain stable for the 7 wavelengths.
Even though the TE$_8$ mode shows the worst performance, its BERs are still below the 7\% FEC threshold of $4.5\times10^{-3}$.
The calculated mode-dependent loss of the system is around 7~dB.
The intensity-averaged impulse response is shown in Fig. \ref{figure3}(c).
Small side lobes are mainly introduced by some weak reflections in the system.
Impulse response broadening due to modal dispersion is negligible due to the short length of the multimode waveguide, which indicates the computational complexity for on-chip MDM optical interconnects can be further optimized and reduced using few equalizer taps.
This demonstration achieves 1.23-Tb/s net capacity and a potential spectral efficiency of 37.2~bit/s/Hz can be reached incorporating with WDM assuming using a 33-GHz wavelength grid.

\section{Conclusions}
In this paper, a record net capacity of 1.23 Tb/s per wavelength over a single SOI multimode waveguide for optical interconnects is experimentally demonstrated by employing on-chip mode- division multiplexing and 11$\times$11 MIMO digital signal processing.
Experimental results show the potential for on-chip MDM technology to sustain ultrahigh bandwidth dataflows for highly dense optical interconnect systems in future data-center and supercomputer network.


\newpage
\printbibliography
\vspace{-4mm}

\end{document}